\documentclass[aps,prb,twocolumn,superscriptaddress,nofootinbib]{revtex4}

\usepackage{amsmath,amssymb,graphicx}
\usepackage{upgreek}
\usepackage{color}
\usepackage[hang,nooneline]{subfigure}
\usepackage[normalem]{ulem}
\usepackage{hyperref}
\hypersetup{colorlinks=true,linkcolor=blue,citecolor=blue,urlcolor=blue}

\definecolor{gray}{rgb}{0.7,0.7,0.7}
\definecolor{orange}{rgb}{1, 0.4, 0}
\definecolor{dgreen}{rgb}{0.0, 0.4, 0.0}
\definecolor{yblue}{rgb}{0.06, 0.3, 0.57}

\def\editr#1{#1}
\def\refer#1{#1}
\def\edstrike#1{}

\graphicspath{{figures/}}

\begin{document}

\title{Interactions and scattering of quantum vortices in a polariton fluid}

\author{Lorenzo Dominici}
\affiliation{CNR NANOTEC, Istituto di Nanotecnologia, Via Monteroni, 73100 Lecce, Italy}
\author{Ricardo Carretero-Gonz\'{a}lez}
\affiliation{Nonlinear Dynamical Systems Group, 
Computational Sciences Research Center, and
Department of Mathematics and Statistics,
San Diego State University, San Diego, California 92182-7720, USA}
\author{Antonio Gianfrate}
\affiliation{CNR NANOTEC, Istituto di Nanotecnologia, Via Monteroni, 73100 Lecce, Italy}
\author{Jes\'{u}s Cuevas-Maraver}
\affiliation{Grupo de F\'{\i}sica No Lineal, Departamento de F\'{\i}sica Aplicada I, Escuela Polit\'ecnica Superior, Universidad de Sevilla, C/ Virgen de \'Africa, 7, 41011-Sevilla, Spain}
\affiliation{Instituto de Matem\'aticas de la Universidad de Sevilla (IMUS). Edificio Celestino Mutis.~Avda.~Reina Mercedes s/n, 41012-Sevilla, Spain}
\author{Augusto S.~Rodrigues}
\affiliation{Departamento de F\'{\i}sica e Astronomia/CF, Faculdade de Ci\^{e}ncias, Universidade do Porto, R.~Campo Alegre, 687 - 4169-007 Porto, Portugal}
\author{Dimitri J.~Frantzeskakis}
\affiliation{Department of Physics, National and Kapodistrian University of Athens, Panepistimiopolis, Zografos, Athens 15784, Greece}
\author{Giovanni Lerario}
\affiliation{CNR NANOTEC, Istituto di Nanotecnologia, Via Monteroni, 73100 Lecce, Italy}
\author{Dario Ballarini}
\affiliation{CNR NANOTEC, Istituto di Nanotecnologia, Via Monteroni, 73100 Lecce, Italy}
\author{Milena De Giorgi}
\affiliation{CNR NANOTEC, Istituto di Nanotecnologia, Via Monteroni, 73100 Lecce, Italy}
\author{Giuseppe Gigli}
\affiliation{CNR NANOTEC, Istituto di Nanotecnologia, Via Monteroni, 73100 Lecce, Italy}
\author{Panayotis G.~Kevrekidis}
\affiliation{Department of Mathematics and Statistics, University of Massachusetts, Amherst, Massachusetts 01003-4515 USA}
\author{Daniele Sanvitto}
\affiliation{CNR NANOTEC, Istituto di Nanotecnologia, Via Monteroni, 73100 Lecce, Italy}
\affiliation{INFN, sezione di Lecce, 73100 Lecce, Italy}

\keywords{quantum vortices $|$ vortex interactions $|$ condensates $|$ polaritons}

\begin{abstract}
\noindent \textbf{Abstract}\\
\editr{
Quantum vortices, the quantized version of classical vortices, play a
prominent role in superfluid and superconductor phase transitions.
However, their exploration at a particle level in open quantum systems
has gained considerable attention only recently.
Here we study vortex pair interactions in a resonant polariton fluid created in a solid-state microcavity.
By tracking the vortices on picosecond time scales,
\refer{we reveal the role of nonlinearity, as well as of
density and phase gradients, in driving their rotational dynamics.}
Such effects are also responsible for the split of composite spin-vortex molecules into elementary half-vortices,
when seeding opposite vorticity between the two spinorial components.
Remarkably, we also observe that vortices placed in close proximity experience
a pull-push scenario leading to unusual scattering-like events that can be described by
a tunable effective potential.
Understanding vortex interactions can be useful in quantum hydrodynamics
and in the development of vortex-based lattices, gyroscopes, and logic devices.
}

\end{abstract}

\maketitle

\noindent \textbf{Introduction}\\
Quantum vortices~\cite{huang_quantum_2015} correspond
to wave field rotations
in systems described by means of complex wavefunctions.
In contrast to the classical case,
the continuity of the phase constrains the phase
circulation (also called phase winding or topological charge $l$)
to be an integer number.
A direct consequence is that the fluid velocity---that in a superfluid is proportional to the phase gradient---decays as $1/r$ away from the vortex core.
Quantum  vortices are commonly observed in a wide range of contexts including
condensates~\cite{stringari,fetter,DarkBook,Lagoudakis2008,Sanvitto2010},
superconductors~\cite{blatter_vortices_1994},
optics~\cite{willner_different_2012,molina-terriza_twisted_2007},
free electron beams~\cite{uchida_generation_2010},
and even in the recently detected gravitational waves
originating from the merging of two spinning black holes~\cite{abbott_observation_2016}.
In particular, quantum vortex configurations
and pairing in condensates are fundamental in
relation to their long-range order coherence, phase transitions, and quantum
turbulence~\cite{BPA,dagvadorj_nonequilibrium_2015,serafini_vortex_2017}.

Nonlinear effects enable the motion of vortices in external
density and phase field gradients~\cite{kivshar_dynamics_1998}.
These interactions result in quantum vortices experiencing two main
driving velocities: one parallel to phase gradients and another
perpendicular to density gradients.
In atomic Bose Einstein condensates (BECs), both the case of cowinding and
that of counterwinding two or few vortices
have been studied~\cite{navarro_dynamics_2013,torres_dynamics_2011,theo,li_dynamics_2016,middelkamp_guiding-center_2011}.
When considering the spinor nature of two-component condensates,
recent works pointed to the possible representation in terms of
vortex molecules~\cite{nitta_vortex_2014, kasamatsu_vortex_2004}
and to the more complex nature of the corresponding
interactions~\cite{pshenichnyuk_pair_2017,tylutki_confinement_2016,kasamatsu_short-range_2016,eto_interaction_2011}.
One driving concept in such theoretical works
is the perpendicular velocity exerted by the vortices on each other~\cite{calderaro_vortex_2017},
resulting in the orbiting/parallel motion of two co-/counter-winding vortex cores, respectively.
Experimental collisional dynamics~\cite{seo_collisional_2016},
were recently induced on the time-scale of seconds
in the case of an antiferromagnetic spinor BEC,
exploiting the counter-rotating orbits of opposite charge vortices.
\edstrike{However, to the best of our knowledge,
the fundamental interactions between
vortices in spinorial condensates have not been studied yet
despite their importance.}
\edstrike{Indeed the} \editr{The} rich phenomenology of vortex dynamics observed in BECs reaches far
out of their specific physical domains, including their role as
cosmological simulators~\cite{Zurek1985,jeff}.
In that context, among the \edstrike{most challenging} proposals for \editr{fundamental} \edstrike{new} physical theories
are schemes describing the quantum vacuum as a special superfluid/BEC
medium~\cite{Huang20161,fedi_superfluid_2016,sbitnev_hydrodynamics_2016_I},
and the elementary particles as quantized vortex excitations on such background~\cite{sbitnev_hydrodynamics_2016_II}.

In this work we use a compact solid-state device to explore
the fundamental nonlinear interactions between
vortices and their dynamics in an exciton polariton BEC~\cite{sanvitto_road_2016,Byrnes2014}.
Semiconductor microcavity polaritons represent a convenient platform
to achieve condensates of
strongly coupled exciton and photon fields~\cite{Byrnes2014,Kasprzak2006},
for the study of two-dimensional (2D) quantum hydrodynamics
and topological excitations~\cite{Lagoudakis2008,Lagoudakis2009,roumpos2,Amo2009,Sanvitto2010,Amo2011,Manni2012} in dissipative and interacting superfluids.
Polariton fluids are hence similar
to nonlinear optics media and atomic BECs,
yet possessing their own peculiarities,
such as Rabi coupling~\cite{Dominici2014},
nonparabolic dispersions (e.g., negative mass)~\cite{gianfrate_superluminal_2018}
and strong nonlinearities~\cite{dominici_real-space_2015}.
One of their assets is the ability to readily
\edstrike{imprint} \refer{generate} a given initial state
(setting velocity, directionality, \editr{spin and orbital angular momentum}, etc.),
therefore providing a full control over the quantum state of the
polariton fluid~\cite{sanvitto_all-optical_2011}.
Polaritons support rich spinorial patterns, in analogy to optical
(multi-frequency or multi-polarization) systems and multi-component BECs~\cite{Kevrekidis2016140,kasamatsu_multi_review_2005}.
\editr{When considering both the spin and orbital angular momentum degrees of freedom,
three basic vortex configurations are most relevant:
(i) the full-vortex, that is composed of phase singularities
with the same orbital charge in both spin populations, 
(ii) the spin-vortex, which, in contrast, is composed of 
opposite windings between the two spin components, and
(iii) the half-vortex, that consists of a unit charge 
coupled to a chargeless configuration.
It is straightforwardly observed that in the full-vortex
case, the resulting polarization is spatially homogeneous,
while for the other two cases a surrounding inhomogeneous 
texture is present.}
Full- and half-integer quantum vortices~\cite{Liew2007,Rubo2007,Voronova2012} have attracted recent interest
since they can be studied both under spontaneous as well as resonant generation,
and offer the possibility to shape vortex-antivortex pair-creation events~\cite{sanvitto_all-optical_2011},
vortex lattices~\cite{Hivet2014,boulier_vortex_2015} and spin-vortex textures~\cite{Manni2013,Liu2015,donati_twist_2016},
together with highly nonlinear dynamics~\cite{dominici_vortex_2015} 
\editr{and, more recently, even Rabi-vortex coupling effects~\cite{dominici_ultrafast_2018}}.
\edstrike{Applications for ultra-sensitive gyroscopes~\cite{Franchetti2012} and
information processing~\cite{Sigurdsson2014} have been theoretically proposed for polariton vortices.}

We \edstrike{imprint} \refer{generate the polariton fluid directly embedding a pair of vortices,} using a resonant pulsed excitation beam with a
modified Laguerre-Gauss (LG) spatial profile.
\refer{Indeed, the resonant excitation results in that the photonic pulse profile is coherently converted
in a polariton fluid, representing its initial condition.}
In this manner
we are able to seed vortices (initially locked
by the resonant photonic \editr{pulse}) at any desired locations \editr{and with a settable intercore distance}.
\editr{The winding direction of the vortex (i.e., $l \equiv \text{ OAM}=\pm1$, orbital angular momentum)
can be set too.
Here we always initialize the fluid with a pair of cowinding vortices, inside a given spin component,
in order to activate self-propelled rotations of the pair itself}.
After the pulsed pump \edstrike{is removed} \refer{completes its injection}, we
track the ensuing vortex dynamics for different pumping \edstrike{rates} \editr{powers} which amounts to
controllably varying the nonlinearity in the system.
When the external pumping is increased---thus the total
population and hence the effective nonlinearity---vortices are able to interact
more strongly.
\editr{
Indeed, this increases the self-rotation effect \edstrike{for cowinding vortices} and,
in the case of a spinorial configuration with opposite orbital charges between the two spin components,
is able to split the spin-vortex states into their
composing half-quantum vortices (i.e., baby-skyrmions\cite{donati_twist_2016}).}
Furthermore, contrary to the case of vortices in atomic, one-component, BECs,
we observe that, for high enough pumping power, \edstrike{after a transient dynamical stage,}
\editr{a radial dynamics is triggered too and the} vortices start to approach each other.
Such \edstrike{unusual} effect grants access to \edstrike{novel} \editr{unexpected} vortex-vortex scattering scenarios.
We deploy a theoretical model to understand why vortices get closer to each
other and conclude that this effect is due to strong spiral density patterns---mediated by 
the intrinsic
excitonic nonlinearity---that channel the vortices to get closer to each other and eventually scatter.
\edstrike{This work can offer hints on the design of polariton vortex devices 
and also sustain novel paradigms
for the interpretation of elementary particles
and of their interactions
in the context of superfluid vacuum theories~\cite{Huang20161}.}
\editr{The understanding and control of vortex-vortex interactions can play a role in
quantum hydrodynamics~\cite{tsubota_quantum_2013},
BECs phase transitions~\cite{dagvadorj_nonequilibrium_2015} and patterns formation~\cite{zhao_pattern_2017}
or superfluid vacuum theories~\cite{Huang20161},
as well as suggesting hints in the design of
vortex lattices shaping~\cite{Hivet2014},
ultra-sensitive gyroscopes~\cite{Franchetti2012} and
information processing polariton devices~\cite{Sigurdsson2014}.}\\

\noindent \textbf{Results}\\
\noindent \textbf{A multicomponent polariton fluid.}
Our physical system consists of four components: the exciton and the
photon fields in both the $\sigma^+$ and $\sigma^-$ \editr{circular} polarizations
\editr{(denoted as $\sigma \equiv \text{ SAM}=\pm1$, spin angular momentum)}.
Both spin polarizations of the photon field can be accessed (measured)
independently in the experiment, while the exciton fields are not
accessible for measurement. However, given that the system is only excited
in one of its two normal modes, the lower polariton branch, the photon
field corresponds to a one-to-one mapping of the exciton field
(apart from a $\uppi$-phase shift) and, in essence, of the polariton field.
In our experiments
(see the Methods section below for more details),
the ultrafast imaging of the quantum fluid over tens of picoseconds,
reveals the in-situ (2+1)D hydrodynamics, where the vortices
can be individually tracked and their full $(x,y,t)$ trajectories retrieved.
This is a significant advantage over atomic BECs, where, typically, only
the density can be monitored (not the phase)
and where only a few snapshots from a given experimental sequence can be obtained~\cite{Freilich1182}.\\

\begin{figure}[htpb]
\centering
\includegraphics[width=1\linewidth]{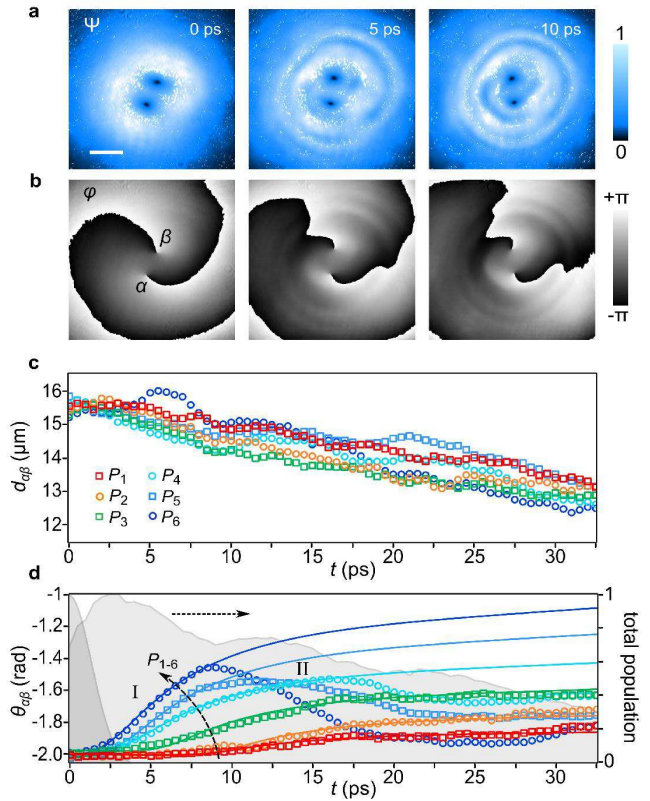}
\caption{{\bf Vortex-vortex \refer{rotational pair} dynamics.}
Polariton (\textbf{a}) amplitude and (\textbf{b}) phase maps
at different times ($t = 0, 5$ and $10$ ps) for the initial total
polariton population $P_6$.
The two phase singularities are \editr{visible as small dark spots} in \textbf{a}
\editr{and labeled as $\alpha$ and $\beta$ in \textbf{b}}.
Intervortex (\textbf{c}) distance and (\textbf{d}) angle for six initial total
polariton populations ($P_{1}$--$P_{6})$.
\editr{The labels I and II
indicate the effective nonlinear effect and associated rotational regime (see text)}.
In \textbf{d} the open symbols are the experimental points and the solid lines represent fitting curves
using the theoretical toy-model introduced
in the \editr{Supplementary Note 6}.
The six pump powers correspond to the following initial
populations: $P_{1-6} \equiv 60\cdot 10^{3}$, $150\cdot 10^{3}$, $300\cdot 10^{3}$,
$0.6\cdot 10^{6}$, $1.0\cdot 10^{6}$, $1.5\cdot 10^{6}$ polaritons.
The background shaded area in \textbf{d} depicts a time-series of
the total polariton population plotted against 
\editr{the normalized $y$-axis to the right,
while the darker half-Gaussian envelope on the left represents the resonant pumping
temporal pulse profile.
\refer{Please note that such time envelope, together with 
the space maps in the first frame ($t=0\text{ ps}$) of \textbf{a},\textbf{b},
jointly represent the full spatio-temporal-phase configuration of the photonic pump pulse,
setting the initial condition of the polariton fluid.}
The time zero ($t=0$) of the dynamics has been chosen at its arrival,
$\sim 2.5 \text{ ps}$ before the maximum of the total emission.
The scale bar to the maps is $20~\upmu\text{m}$.}
}
\label{fig:FIG1}
\end{figure}

\noindent \textbf{Vortex dynamics.}
The spatial structure of the vortices is reported in Fig.~\ref{fig:FIG1}\text{a},\text{b},
where the left panels show the photonic emission from the initial state of the polariton fluid
\refer{(which bears the same spatial profile as the resonant photonic pump)}.
The pumping density and phase distribution is the same for all the
different pulse powers used here---that in turn correspond to the initial total
polariton populations denoted $P_{1-6}$,
see Fig.~\ref{fig:FIG1}---and, more importantly, the phase winding is
the same in \editr{both} \edstrike{the} spinorial components \editr{(i.e., we are initializing a pair of full-vortices)}.
We note that we
only display in Fig.~\ref{fig:FIG1}
the $\sigma^+$ spin polarization photon field as
the $\sigma^-$ field is perfectly synchronized and follows, indistinguishably, the dynamics of
the $\sigma^+$ field 
(\editr{for more details please see below the section on Spinorial vortex configurations}\edstrike{for instance Fig.~\ref{fig:FIG4}\text{a--c}}).
The phase map allows for the precise determination of vortex locations
(small black spots in the amplitude maps \editr{and labeled as $\alpha$ and $\beta$ here and in the following})
\edstrike{, inside any of the two spin populations (denoted as $\sigma \equiv \text{ SAM}=\pm1$,
spin angular momentum)}.
It is evident that the total topological charge
(i.e., \edstrike{the orbital angular momentum per polariton, $l \equiv \text{ OAM}$} \editr{$l=2$}),
is composed of two separated cores in the central
region.
The initial vortex separation, which can be controlled upon proper tuning of the
optical \editr{pulse-shaping device (see Methods section below)} \edstrike{phase-shaping}, is $\sim16~\upmu\text{m}$.
In Fig.~\ref{fig:FIG1}\text{a,b} we show three snapshots of the fluid  at time intervals of $5~\text{ps}$,
for the case with the largest density ($P_6$).
Note that the vortices
rotate around the center of the configuration
and approximately maintain the same mutual distance,
while some circular
ripples are induced in the density.
It is also relevant to mention that the initial size of vortex cores (as seeded by the pump)
is about three times larger than
the healing length $\xi$
expected from 
the initial polariton density ($\xi \sim 4.5~\upmu\text{m}$ for $P_1$ at $2.5~\text{ps}$).
However, at very long times, when the polariton population decreases significantly
(population at $30~\text{ps}$ is about $30\%$ of the maximum population), 
$\xi$ should increase by less than a factor of two.
Nevertheless, since the healing length represents a \editr{minimal} value for the vortex core size,
we can exclude any substantial effect of $\xi$ on the \refer{cores size during the} observed dynamics, 
for any of the six different power regimes
\refer{(as a matter of fact, the core size stability during the relevant time ranges
can be also noted in all the amplitude maps shown here and in the following sections).
On the other hand, this also confirms the strong out-of-equilibrium nature of the initial configuration
and suggests that an ensuing relaxation is expected to affect the fluid reshaping and, thus,
take part in mediating the inter-vortex rotational and radial effects described in the following.
For this reason, we describe the observed dynamics in terms of what could be considered
as effective vortex-pair interactions.
A possible alternative, genuine, quasi-equilibrium regime could also be studied upon reaching 
thermalization (where core sizes would relax to their natural healing length) in, for example,
ultra-high quality samples with lifetimes of hundreds of picoseconds~\cite{caputo_topological_2017}.}\\

The vortex-vortex dynamics are summarized in the time plots
in Fig.~\ref{fig:FIG1}\text{c},\text{d} which
show, respectively, the intervortex distance $d_{\alpha\beta}(t)$ and
angle $\theta_{\alpha\beta}(t)$ in the $\Delta t = 0 \text{ -- } 32.5~\text{ps}$ time range
\refer{(please note that here and in the following,
the arrival of the photonic pulse
has been chosen as the time zero of the dynamics, 
while the maximum polariton population is reached after $2.5 \text{ ps}$)}.
The cores separation in Fig.~\ref{fig:FIG1}\text{c} slightly decreases during the whole dynamics,
with approximately the same constant speed for all the different initial densities
 (that is an average $v \sim 0.04~\upmu \text{m} \cdot \text{ps}^{-1}$ for each of the two vortices).
We ascribe this continuous slow vortex approach to the presence of an
inward phase gradient in the pump beam
as it can be clearly \edstrike{observed from} \editr{deduced from the presence of} the spiral phase patterns in Fig.~\ref{fig:FIG1}\text{b}
\edstrike{We label this slow inward drift as driving action I} 
\editr{(see Supplementary Note 3 for more details)}.
Such a gradient represents an external linear drive that
was set upon fine tuning of the optical focusing of the pump,
in order to weakly push the vortices towards each other.\\

\noindent \editr{\textbf{Nonlinear rotational effects.}}
The nonlinear rotation of the vortices is evident from
Fig.~\ref{fig:FIG1}\text{d} (such \edstrike{driving action} \editr{nonlinear effect I, with its associated rotational regime, are} manifest since the early times of the dynamics,
the precise duration depending on power).
It is crucial to note that this rotational effect is generated
by the circular superfluid currents (proportional to the \editr{azimuthal phase gradients})
independently generated by each vortex on the location of its partner.
Furthermore, we observe that the rate of rotation of the vortices
increases for stronger pump powers. This is \editr{equivalent} to 
the case in atomic BECs where the vortex interactions
increase with the strength of the nonlinearity in the system.
\refer{This can be physically interpreted by noting that 
increasing the density also increases the speed 
of sound, and thus one would expect faster 
motion of the vortices as they interact.}
In our case, as the pump power is increased, the polariton population increases
and\edstrike{, thus,} \editr{so does} the effective nonlinearity---proportional to the exciton
density [see Eq.~(\ref{eq:coupled_GPEs_1})].
At a first order approximation, this action is expected to depend on the instantaneous
and local density, which rises during
the pump pulse arrival and then exponentially decays due to
dissipation associated with polariton lifetime (mainly due to photon emission).
The overall effect of this rise and fall of the effective nonlinearity is expected
to induce a fast rotation followed by a slow deceleration in time,
qualitatively resulting in an overall sigmoid shape of the $\theta_{\alpha\beta}(t)$ curves.
However, such a simplified scheme is able to capture the vortex dynamics
only for short times,
as shown by the solid lines in Fig.~\ref{fig:FIG1}\text{d} which are fitting
curves using the theoretical toy-model
introduced in \editr{the Supplementary Note 6}.
In the experiments, we observe
the presence of an additional counter-rotating effect at later times for the largest powers
(\edstrike{driving action} \editr{rotational effect and regime II}, manifesting at $t = 17, 12$ and $8~\text{ps}$ for $P_{4,5,6}$, respectively)
which leads to a reversal of the rotation (rather than simply to its saturation).
Here, the nonlinear reshaping of the fluid results in the formation of circular density ripples
whose radial gradient represents an additional azimuthal drive on vortex motion.
As the vortices ride on the \editr{inside} of 
these (circular) radial ripples\editr{---previously reported~\cite{Liew2007,dominici_real-space_2015,dominici_vortex_2015}
and studied from a bifurcation
perspective as ring dark (gray) solitons~\cite{Rodrigues2014}}---they provide a sharp
\editr{negative} radial density gradient that is responsible for the vortices rotating
in a \editr{clockwise} direction that overcomes the mutual vortex-vortex
counter-clockwise motion.
Our numerical results, implementing the generalized open-dissipative
Gross-Pitaevskii model described in the Methods section \editr{(also see Supplementary Note 1)},
do produce the radial ripples, however
they are generated further away from the center and, therefore, they do not
affect the vortex rotation (i.e., the model does not precisely reproduce the intensity or the
spatial location of the \edstrike{action} \editr{rotational effect} II).
Nonetheless, we have checked that by changing the strength of the nonlinear
interaction terms 
and/or the size of the pump spot width, and the initial location of the vortices,
it is possible to qualitatively reproduce this rotation reversal 
\editr{(see Supplementary Note 2)}.
Furthermore, our numerical simulations also
reproduce the main experimental observations,
including the slow vortex approach (\editr{inward drift induced by} the pump radial phase),
the increase of the rate of rotation as the nonlinearity
increases (\edstrike{driving action} \editr{effect and regime} I), and also corroborates the sigmoid saturation
which is present in the experiment for low enough powers \editr{(i.e., the $P_{1,2,3}$ cases)}.\\

\begin{figure*}[htbp]
\centering
\includegraphics[width=1\linewidth]{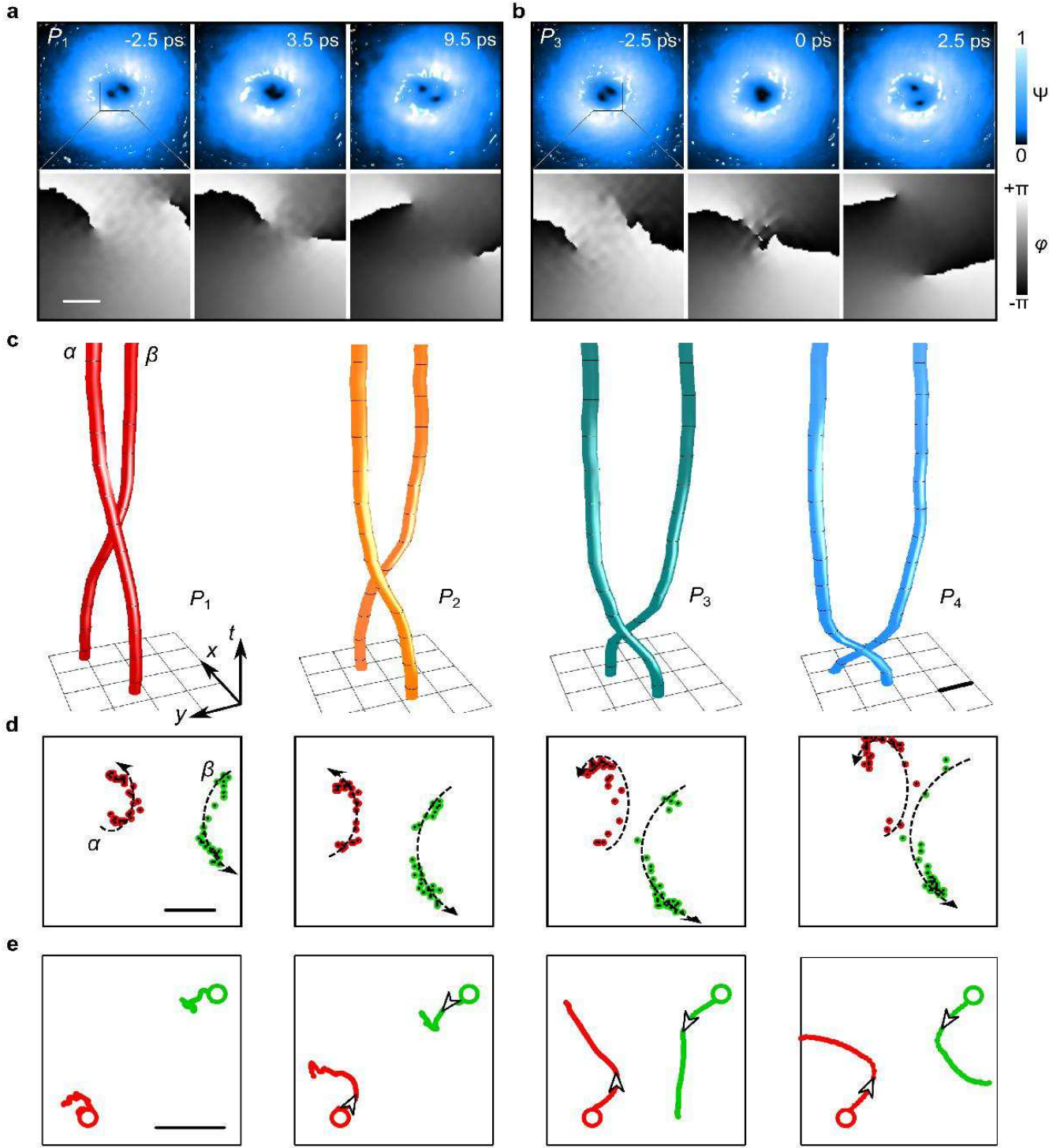}
\caption{{\bf Vortex scattering.}
\textbf{a},\textbf{b}, Amplitude and phase maps of the polariton fluid at three time frames
for a low ($P_1$, \textbf{a}) and a larger ($P_3$, \textbf{b}) initial density.
The phase maps correspond to the small square area of the density maps.
\textbf{c}, The $(x,y,t)$ vortex trajectories (time range $\Delta t=-2.5 \text{ -- } 9.5\text{ ps}$,
step $\delta t=0.5\text{ ps}$) for four increasing initial populations $P_{1-4}$
(the number of polaritons are reported in the caption to Fig.~\ref{fig:FIG3}).
\textbf{d}, Vortex scattering maps for the same four powers,
in the $\Delta t=-2.5 \text{ -- } 13.5\text{ ps}$ timespan.
The two arc arrows are guides for the eye
to help understand their movement over time.
The average and top speeds of the phase singularities are on the order of $\leq1~\upmu\text{m} \cdot \text{ps}^{-1}$
and $\sim10~\upmu\text{m} \cdot \text{ps}^{-1}$, respectively.
\textbf{e}, [arranged in increasing pump power] show the numerical
vortex orbits leading to scattering for large powers. Initial
positions are depicted with circles.
\editr{All the scale bars are $5~\upmu\text{m}$ while the small square box in the top rows of \textbf{a},\textbf{b} is $20\times20~\upmu\text{m}^2$.}
}
\label{fig:FIG2}
\end{figure*}

\begin{figure*}[htbp]
\centering
\includegraphics[width=1.00\linewidth]{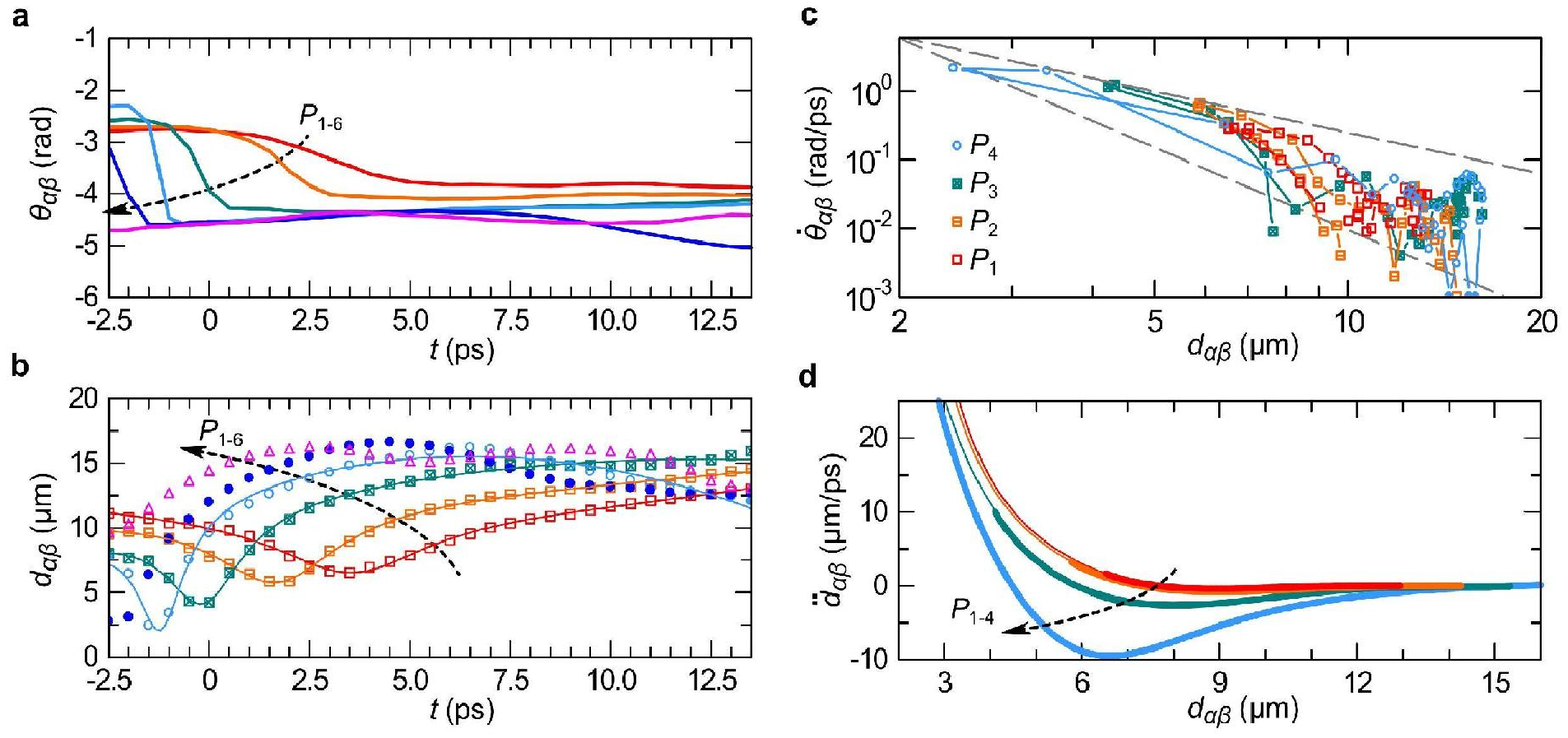}
\caption{{\bf Azimuthal and radial scattering dynamics.}
Evolution of the intervortex (\textbf{a}) angle and (\textbf{b}) distance for different pulse powers.
In \textbf{b} the dots are the experimental points and the solid lines
correspond to fits using a parabola minus a Lorentzian curve 
$d(t) = d_0 + m t - a t^{2} - b/[(t-t_{sc})^{2}+\sigma_{sc}^{2}]$
(see relevant discussion in the text).
The six initial total populations correspond to $P_{1-6} \equiv 48\cdot 10^{3}$,
$96\cdot 10^{3}$, $230\cdot 10^{3}$, $460\cdot 10^{3}$, $0.96\cdot 10^{6}$, $2.4\cdot 10^{6}$ polaritons.
$P_{1-4}$ correspond to the dynamics reported in Fig.~\ref{fig:FIG2}.
\textbf{c}, Intervortex angular velocity vs.~distance in the same timespan as above in a log-log representation.
The two dashed lines represent $1/d^2$ and $1/d^4$ slopes.
\textbf{d}, Radial acceleration $\ddot d$ vs.~distance $d$.
The thick lines are obtained upon differentiating the fitted $d_{\alpha\beta}(t)$ curves in \textbf{b} and plotted as a function of $d_{\alpha\beta}$,
while the thin lines, added for guidance, represent their $\ddot d(d)$ fits using a
negative power law of $d$ plus a Lorentzian well.
}
\label{fig:FIG3}
\end{figure*}

\noindent \textbf{Scattering-like events.}
We explore the scattering between vortices \editr{after} initially placing the
two cores closer to each other. \editr{For this set of
scattering-like experiments, we only seed vortices in the
$\sigma^+$ spinorial component while using a plain (vortex-less)
Gaussian profile in the $\sigma^-$ component. We use a relative
polariton population in the $\sigma^-$ equivalent to $\sim1/4$
of the polariton population in the $\sigma^+$ component.
We have checked in our model that changes in the relative populations
between the $\sigma^+$ and $\sigma^-$ spinorial components do not
qualitatively change the results hereby presented.
In the following we only discuss the relevant $\sigma^+$ field.}
The corresponding experimental dynamics are
shown in Fig.~\ref{fig:FIG2},
where the initial core separation is $d_{\alpha\beta}\sim 10~\upmu\text{m}$.
The first two rows in the figure show the amplitude and phase maps at the initial time
and successive instants, for two powers $P_1$ (Fig.~\ref{fig:FIG2}\text{a}) and
$P_3$ (Fig.~\ref{fig:FIG2}\text{b}). In both cases there is a given time frame
(respectively $t=3.5\text{ ps}$ and $t=0\text{ ps}$\edstrike{, see middle panels})
at which the two vortex cores appear to merge and then separate again (see right panels).
When the vortices reach their closest proximity (comparable with their core radius),
they cannot be clearly resolved apart in the amplitude maps.
Given the access to the phase maps, where the phase singularities can be pinpointed
with pixel resolution, the dynamics of the point-like entities can be tracked even
when the vortex cores are nearly overlapping with each other
(see also \editr{Supplementary Movies 1--4}, reporting the amplitude and phase
for the first four powers $P_{1,2,3,4}$, respectively).
The associated $(x,y,t)$ trajectories extracted from the phase maps are reported
in the panels of Fig.~\ref{fig:FIG2}\text{c},
for the time range $\Delta t=-2.5 \text{ -- } 9.5\text{ ps}$.
The time-space vortex filaments highlight the approach and bounce-back
of the two cores that, after the scattering, emerge rotated compared
to their initial locations.
The deformation of the vortex strings in the $(x,y,t)$ domain
is related to the nonlinear energy stored and released by the fluid
during the scattering. Nevertheless, these coherent structures
robustly emerge (as individual entities)
from the scattering events.

The vortex-vortex collisions are mapped in Fig.~\ref{fig:FIG2}\text{d}, as the $(x,y)$
trajectories for the two vortices\edstrike{ labeled as $\alpha$ and $\beta$}.
Upon larger initial densities,
the phase singularities reach a more intimate proximity,
confirming a nonlinear scattering-like process.
Trajectories from the numerical simulations are reported in Fig.~\ref{fig:FIG2}\text{e},
and qualitatively reproduce the experiments.
They show how the phase singularities,
slightly wandering at low power, go through stronger
scattering paths upon increasing the population.
We point out that in the numerical model
the scattering-like events are also observed
when starting with an outward phase gradient of the resonant pump,
 when increasing the density (\editr{see Supplementary Note 3}).
This is a further confirmation that the scattering events are an inherent nonlinear effect,
independent from the external action of the initial pump gradient.\\

The collisional features are recognizable in the time plots for
the intervortex angle and distance in Fig.~\ref{fig:FIG3}\text{a},\text{b}, respectively.
The angle $\theta_{\alpha\beta}(t)$ remains approximately
constant before suddenly suffering a sharp change and then settling again
(sigmoid feature).
The intervortex distance $d_{\alpha\beta}(t)$ reaches a minimum for a time in
close correspondence to the inflection point of the angle $\theta_{\alpha\beta}(t)$ curve,
hence at the maximum of the angular speed, before the two vortices bounce back.
We performed a fitting of the $d_{\alpha\beta}(t)$ curves, to retrieve an empirical trend
for the collisional events upon larger excitation density.
The scattering time $t_{sc}$ and time width $\sigma_{sc}$ parameters correspond to:
3.6, 1.7, $-0.2$, $-1.2$ \text{ ps} and 4.7, 3.8, 2.3, 1.4 \text{ ps}, respectively,
for the $P_{1-4}$ cases.
The results for this set of experiments, and the corresponding modeling,
highlight that earlier (and faster) scattering events are associated to larger powers.
The rotational component during the scattering events is outlined
in Fig.~\ref{fig:FIG3}\text{c},
showing the angular velocity as a function of the separation.
All the $P_{1-4}$ curves represent a decreasing trend in the
$\dot{\theta}(d)$ dependence,
and lie between $1/d^2$ and $1/d^4$ power laws.
It is important to contrast this observed trend in the context of point-vortices
in superfluid BECs.
Since the tangential superfluid velocity in a BEC vortex
is inversely proportional to the distance from the core,
the angular velocity of a vortex pair rotating under the mutual azimuthal interaction~\cite{Jackson1998,DarkBook,Kevrekidis_MPLB_2004}
is proportional to $1/d^2$.
However, in our system $\dot\theta(d)$ seems to decay faster.
This is attributed to the fact that the derivation of $\dot\theta\propto 1/d^2$
assumes an effectively constant density background, while in the polariton
case there is an exponentially decreasing density with time (similar
results are obtained with the numerical model, \editr{see Supplementary Note 5}).\\
%

\begin{figure*}[htbp]
\centering
\includegraphics[width=1.00\linewidth]{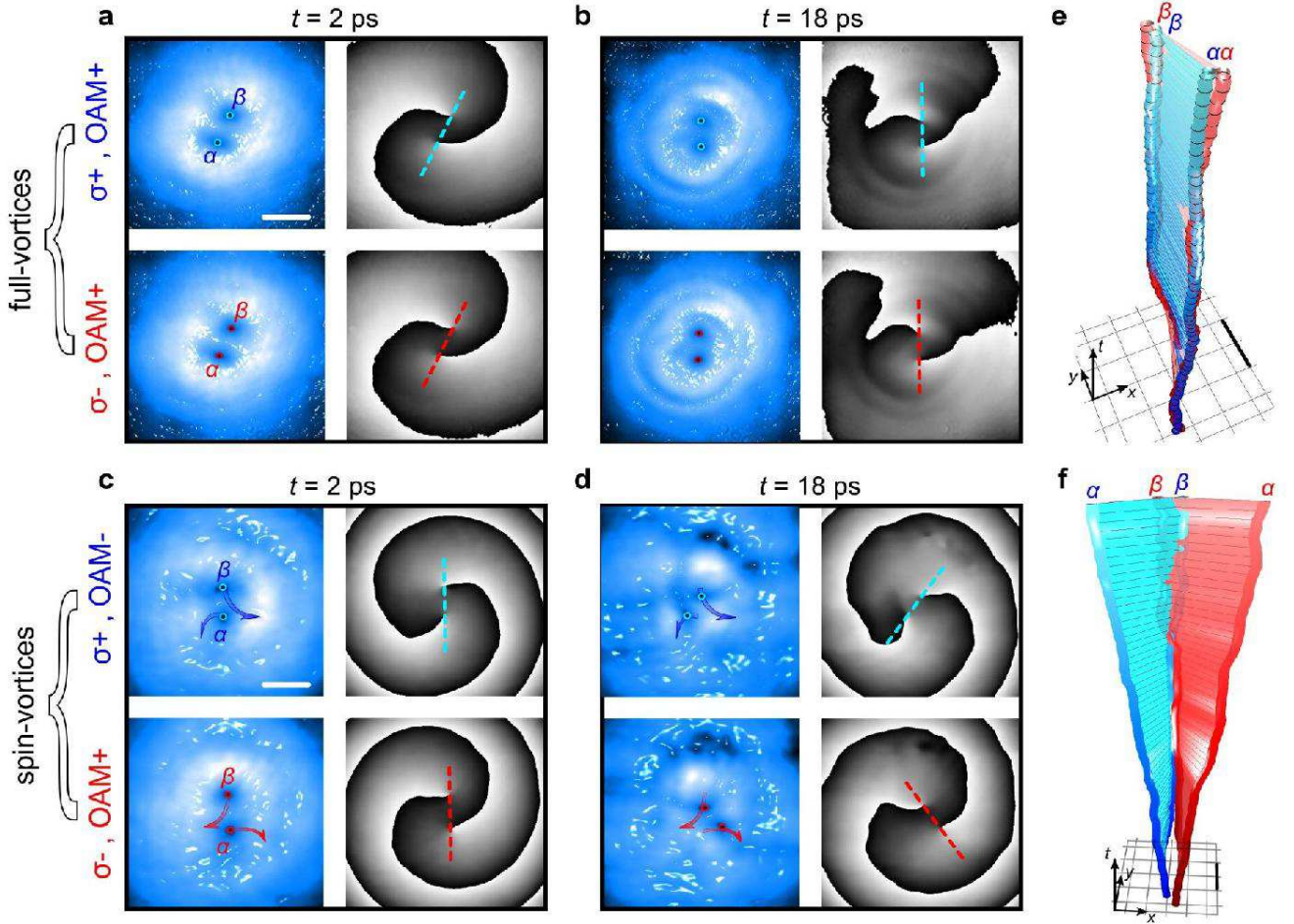}
\caption{{\bf Nonlinear rotation of spinorial vortex pair configurations.}
\textbf{a},\textbf{b}, Amplitude and phase maps of the polariton fluid,
when starting with two \editr{full-}vortices (same \editr{OAM} in the two opposite spin populations),
at \editr{initial and later} time frames.
\textbf{c},\textbf{d}, Initialization with two \editr{spin-}vortices (opposite \editr{OAM} 
between the two spin populations), amplitude and phase maps of the polariton fluid at different time frames.
Movements of the phase singularities are tracked as blue and red dots
in the amplitude maps, while the solid \editr{arrows in \textbf{c},\textbf{d}} represent the two vortex trajectories
during the whole dynamics ($\Delta t = 0 \text{ -- } 35 \text{ ps}$).
\editr{The initial total population is $P \sim 450\cdot 10^{3}$ (i.e., intermediate between the $P_3$ and $P_4$ cases of Fig.~\ref{fig:FIG1}).} 
\textbf{e},\textbf{f}, Vortex $(x,y,t)$ trajectories for (\textbf{e}, $\Delta t = -4 \text{ -- } 50 \text{ ps}$) cowinding and
(\textbf{f}, $\Delta t = 0 \text{ -- } 31 \text{ ps}$) counterwinding cases.
\editr{The scale bars are $25~\upmu\text{m}$ in the space maps (\textbf{a},\textbf{c}) and $10~\upmu\text{m}$ in the $(x,y,t)$ plots (\textbf{e},\textbf{f}).
The color code of labels, arrows and vortex tubes indicates the right (blue, $\sigma^+$) and left (red, $\sigma^-$) spin components.
The dashed lines in the phase maps help highlight the opposite rotations of the vortex pairs depending on their OAM direction.}
\refer{Please note that the amplitude and phase maps at  $t = 2\text{ ps}$ (\textbf{a,c}) represent both the photonic pump profiles 
(for the two spinorial configurations) and the initial condition of the polariton fluid.}
}
\label{fig:FIG4}
\end{figure*}

\noindent \editr{\textbf{Extracting an effective potential.}}
Finally, we show the radial acceleration during
the scattering events in Fig.~\ref{fig:FIG3}\text{d}.
This plot helps to interpret the
collisional dynamics as driven by an effective radial pull-push.
For relatively large distances, the effective force is approximately
zero leading to circular-like motion;
while, at shorter ranges, the effective force
acquires a negative component (and stronger for higher densities)
and thus induces the vortices to get closer to each other.
It is important to mention that
the curves depicted in Fig.~\ref{fig:FIG3}\text{d}
are drawn under non-equilibrium conditions and
cannot be straightforwardly assigned to a genuine pairwise potential between the vortices.
Nonetheless, the results suggest that it is possible to induce
two cowinding vortices
to get closer to each other and modulate (increase) the rate of approach
upon increasing the condensate's density.
Numerical simulations of our model of
Eqs.~(\ref{eq:coupled_GPEs_1})-(\ref{eq:coupled_GPEs_2}) suggest that,
while the polariton vortices drag each other in a mutual circular dance,
as standard cowinding vortices do,
they also induce local spiral density patterns
self-channeling their approach and scattering.
A cleaner depiction of this self-channeling density spiral
is presented in \editr{Supplementary Note 4}
(see also \editr{Supplementary Movie 5}).
Parametric explorations within the numerical model allow us to conclude
that the vortex approach is mediated by a combination
of the photonic kinetic energy term, the Rabi coupling between
the photonic and excitonic components,
and the intrinsic excitonic nonlinearity.
The induced \editr{attraction} represents a novel fundamental effect to be
used at the basis of more complex
quantum hydrodynamics and turbulence scenarios
as well as in the nonlinear shaping of multi-component vortex lattices.\\

\begin{figure*}[htbp]
\centering
\includegraphics[width=1.00\linewidth]{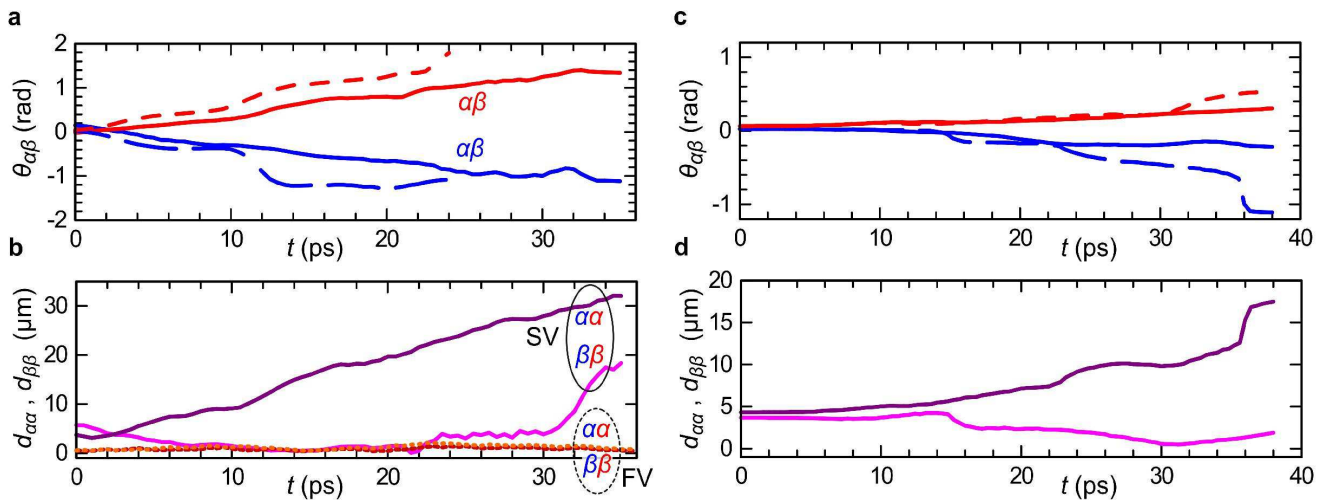}
\caption{{\bf Rotational split of composite spin-vortices.}
\textbf{a}, Evolution of the intervortex angle \editr{$\theta_{\alpha\beta}$} in the \editr{full} timespan
(solid blue and red lines \editr{for the $\sigma^+$ and $\sigma^-$ spin components, respectively) in the spin-vortex} case
of Fig.~\ref{fig:FIG4}\textbf{c},\textbf{d} and \textbf{f}.
A larger power is reported too (dashed blue and red lines).
\textbf{b}, Interspin distances $d_{\alpha\alpha}^{\pm}$ and $d_{\beta\beta}^{\pm}$
between the \editr{corresponding cores} in the two spin populations, for both \editr{counterwinding (SV, spin-vortex, solid lines) 
and cowinding (FV, full-vortex, dashed lines) configurations}.
\textbf{c}, Numerical intervortex angle (intraspin) 
\editr{$\theta_{\alpha\beta}$}
and \textbf{d}, intervortex
distances (interspin) $d_{\alpha\alpha}^{\pm}$ and $d_{\beta\beta}^{\pm}$ for the
counterwinding case \editr{(spin-vortex)}, corresponding to the experimental panels \textbf{a} and \textbf{b}, respectively.
\editr{The $\alpha$ and $\beta$ labels indicate the spatially separated, cowinding, vortex cores inside each of the two spin populations.}
}
\label{fig:FIG5}
\end{figure*}

\noindent \editr{\textbf{Spinorial vortex configurations.}}
\editr{The phase singularities in each of the two spin populations
represent elementary point-like particles with two associated quantum numbers,
the SAM and OAM (e.g., ${\alpha}^{\text{SAM}}_{\text{OAM}}$,
where $\alpha$ represents the specific vortex core and SAM and OAM 
its unitary spin and orbital angular momentum charge).
Depending on the direction of the quantum numbers,
different vortex combinations are possible.}
Here \edstrike{we} \editr{it is relevant to} explore
the effect of the mutual azimuthal thrust on the core dynamics
when seeding composite vortices
relying on same, as well as opposite,
charges between the different spin components
\editr{(i.e., when starting with full- or spin-vortex pairs, respectively)}.
In contrast to the results presented in Fig.~\ref{fig:FIG1}
where both $\sigma^+$ and $\sigma^-$ spin polarizations
remained synchronized during evolution, here the different interactions
inside each of the two spin polarizations result in the two components
evolving independently. We therefore need to individually measure
and display the evolution in each spin component.
The results are reported in Fig.~\ref{fig:FIG4} and validate that the
rotation effect and its direction are due to the phase drive of the
two vortex currents \edstrike{on each other} \refer{which are present} inside the \editr{same} spinorial
component. 
Indeed, in the cowinding case, the rotation is in the same
direction for both the $\sigma^+$ and $\sigma^-$ vortex doublets,
as shown in Fig.~\ref{fig:FIG4}\text{a,b}.
The initial configuration
is almost identical in both spin populations
albeit with a small deviation of the vortex positions
between components.
Despite this small initial deviation, the vortices across components stay close
to each other and the overall dynamics between components stays
synchronized, due to the same
\edstrike{rotational interactions} \refer{nonlinear effects} (\editr{rotational regime I})
existing in both spin populations.
It is also possible that the weak attractive inter-spin interactions
help in stabilizing against small differential disorder between the spin
populations~\cite{dominici_vortex_2015}.
In contrast, for the counterwinding case, the observed trajectories
(see blue and red orbits in Fig.~\ref{fig:FIG4}\text{c,d})
are (left-to-right) mirrored across the components.
The (2+1)D vortex lines for both co- and counter-rotating cases
are reported in Fig.~\ref{fig:FIG4}\text{e,f}, respectively.
\editr{In such panels we emphasize the co- and counter-rotations}
by using two lines (blue and red, at any time frame) linking the $\alpha$ and $\beta$ cores
inside any of the two spin populations [and with the lines drawing two sheets in the $(x,y,t)$ domain].\\

In the time plots of Fig.~\ref{fig:FIG5}\text{a} we confirm the opposite rotations
(as solid blue and red line) in time
$\theta_{\alpha\beta}(t)$, and the faster rotation when increasing power (as dashed lines).
The separation of the corresponding $\alpha$ and $\beta$ \edstrike{charges} \editr{cores} between the two spins
is reported in Fig.~\ref{fig:FIG5}\text{b} (\editr{please note that such} distances are labeled as $d_{\alpha\alpha}^{\pm}$
and $d_{\beta\beta}^{\pm}$).
The resulting counter-rotation of the doublets
leads to the separation of at least one interspin couple, for which the distance increases
approximately linearly (see $\alpha\alpha$, solid purple line).
Due to slight asymmetries in the initial conditions (not a perfect alignment/tuning of the phase shaping),
the second couple separation actually decreases
(although, for later times,
we see it separating as well,
see $\beta\beta$, solid magenta line).
These observations are also exhibited by the
numerical simulations,
that use the experimental profiles as an initial condition.
In fact, the numerical plots of the intervortex angles
and distance in Fig.~\ref{fig:FIG5}\text{c},\text{d} qualitatively reproduce those of Fig.~\ref{fig:FIG5}\text{a},\text{b}.
In contrast, the cowinding and co-rotating doublets preserve
the initial overlapping for the corresponding cores between the spins (the interspin
cores distance is $\sim1~\upmu\text{m}$ during the whole dynamics, see dashed
light and dark orange lines in Fig.~\ref{fig:FIG5}\text{b}).

In summary, the above set of experiments show that
the vortex-vortex interactions within the same spin component
dominate the dynamics,
while the interactions across the components are weaker and, thus,
are not essential to understand the rotational dynamics.
In fact,
further numerical tests, including cases varying the
strength, and even eliminating, the spin-orbit coupling yielded
almost identical results.
Nonetheless, the two independent intra-spin rotational drives
can affect the overall resulting spinorial state.
For instance, topological charges \edstrike{imprinted} \refer{seeded} as
full-vortices keep rotating jointly. 
In contrast, when \edstrike{imprinted} \refer{seeded} as spin-vortices
they dissociate due to the effect of the opposite nonlinear rotation.
Therefore, the two couples get split into four \editr{half-vortices} \edstrike{baby-skyrmions}
\cite{Rubo2007,Manni2013,donati_twist_2016,dominici_vortex_2015}.\\

\noindent \editr{\textbf{Discussion}}\\
We have studied the external and internal \edstrike{driving actions}
\editr{effects governing the dynamics}
of two quantum vortices
in a nonlinear and out-of-equilibrium 2D polariton fluid
\editr{and discussed these in terms of effective vortex interactions.}
The results show that stronger density regimes
enhance the \edstrike{vortex-vortex} \refer{effective pair} interactions,
accelerating or even reversing the mutual rotation\editr{al} effect
and giving rise to unexpected radial dynamics.
Indeed, nonlinearity fuels the azimuthal phase drive between vortices
resulting in an increase of the speed of the circular motion for same-charge vortices.
Strikingly, at short range,
the intrinsic excitonic nonlinearity induces
an effective radial thrust
that compels the cores to approach and bounce back from each other.
We exploit this feature
to demonstrate \edstrike{novel} \editr{unusual} vortex-vortex scattering\refer{-like} events.
These represent an original scenario that could be further investigated in other multicomponent superfluids,
and would be particularly interesting in the nonlinear shaping of vortex or vortex-antivortex lattices.
\editr{More fundamental, yet intriguing, could be the case of studying the analogue dynamics when seeding opposite-charge vortex pairs (i.e., vortex-antivortex) 
and look for the presence and nature of a similar effective radial thrust, as well as its 
possible influence on pair-annihilation and nucleation events.}

\edstrike{The phase singularities in each of the two spinorial populations
represent elementary point-like particles with two associated quantum numbers,
the SAM and OAM (e.g., ${\alpha}^{\text{SAM}}_{\text{OAM}}$).
Depending on the direction of the quantum numbers,
different vortex combinations are possible.
Indeed we} \editr{We} further study the effects of 
seeding opposite-charge couples
between the two spinorial components. 
As a result, the vortex-vortex
doublets' rotational direction in\editr{side} each of the two spin is opposite,
 which in turn
splits these composite spin-vortices into \editr{half-vortices}\edstrike{ baby-skyrmions}.
These structures, consisting of a unitary charge coupled to
a chargeless configuration,
are relevant in a wide range of fields
within optical, nonlinear, atomic or high energy physics,
where they are also known as Poincar\'{e} beams, vortex-bright solitons,
 \edstrike{half-vortices}\editr{ baby-skyrmions} or filled-core vortices~\cite{Cardano2013,DarkBook,Kevrekidis2016140}.
\editr{
An interesting extension for further exploration would be to study,
e.g., the interplay between two possible competing actions:
the rotational split of the spin-vortices
(due to the intra-spin polariton nonlinearities)
and their stabilization upon the locking of the phase singularities
(by the weaker inter-spin attractions).
The splitting dynamics could  also be investigated for the case of 
opposite charge vortices which tend to move parallel to each other.}

\bigskip

\noindent \textbf{Methods}\\
\small{
\noindent \textbf{Experimental methods.}
\edstrike{The polariton device and the main setup of the experiments are similar
to the ones used in Refs.~\cite{Dominici2014,dominici_real-space_2015,dominici_vortex_2015,gianfrate_superluminal_2018}.
More specifically, w} 
\editr{W}e used a microcavity sample
with an AlGaAs 2$\lambda$ optical thickness and three 8 nm
In$_{0.04}$Ga$_{0.96}$As quantum wells (QWs), placed in the antinodes of the
cavity mode~\cite{Dominici2014,dominici_real-space_2015,dominici_vortex_2015,gianfrate_superluminal_2018}. 
The multilayer mirrors embedding the
cavity consist of AlAs/GaAs layers, with an overall photonic
quality factor of $Q=14000$ and a resulting effective lifetime for the (lower)
polariton fluid of $\sim$ 25 ps, at normal incidence \edstrike{($k=0$)}
and at a temperature of 10~K.
The quality factor of the device is associated to a photon lifetime of $\sim 5\text{--}6\text{ ps}$,
resulting in a polariton lifetime of $\sim 10\text{--}12\text{ ps}$.
However, time-delayed reflections back from the substrate edge
help in sustaining the polariton fluid population for a longer time,
resulting in the longer effective lifetime stated above.
The substrate optical thickness of 1.5 mm
corresponds to a 10 ps time distance of the reflected echos~\cite{dominici_real-space_2015}
\edstrike{(see also Ref.~\onlinecite{dominici_real-space_2015})}.
We performed the experiments in a defects-clean region of the sample,
usually a $100~\upmu\text{m}$ wide square area contained between four line dislocations.
The excitation and reference beams are laser pulses with 80 MHz repetition
rate and $4~\text{ps}$ time width ($0.3~\text{nm}$ bandwidth).
The picosecond excitation and its tuning allow 
the exclusive initialization of the lower polariton mode,
centered at $\sim836.5~\text{nm}$,
while the upper mode that is 3 nm above
(Rabi splitting of $5.4~\text{meV}$)
 is not excited.\\

\noindent \editr{\textbf{Vortex generation.}}
Double optical vortices are \edstrike{imprinted} \refer{seeded} into the system by
phase-shaping the plain-Gaussian LG$_{00}$ laser pulse,
into a modified Laguerre-Gaussian LG$_{0\pm2}$ state
upon passage on the $q$-plate, which is a patterned liquid crystal phase
retarder~\cite{dominici_vortex_2015,Cardano2013,Marrucci2006}.
Indeed the initial splitting of the two unitary charges composing the LG$_{0\pm2}$ state
can be set upon proper tuning of the $q$-plate.
Power control is set upon the use of a $\lambda/2$ plate and of a linear polarizer before the $q$-plate.
The photonic vortex is focused at normal incidence on the sample by means of a $10~\text{cm}$ aspherical lens,
to resonantly excite the polariton fluid.
Polarization control is implemented by means of $\lambda/2$ and $\lambda/4$ plates and of linear polarizers,
prior to and after the $q$-plate in order to achieve the desired combinations
of topological states between the two spin populations.\\

\noindent \editr{\textbf{Ultrafast holographic imaging.}}
The dynamical imaging of the polariton fluid is based on the ultrafast
implementation of the so-called off-axis digital
holography~\cite{Dominici2014,dominici_real-space_2015,dominici_vortex_2015,gianfrate_superluminal_2018,Anton2012,Nardin2010,Schnars_book2005a}.
The sample emission is let interfere with a delayed
and coherent reference beam, which is an expanded twin copy of the
Gaussian excitation beam, able to provide amplitude and phase homogeneous fronts.
The emission and reference beams are sent on the CCD (charge coupled device) camera with a mutual
angle of inclination \edstrike{($\Delta k$)}.
This allows to obtain interferograms which are associated
only to the time portion of the emission synchronous
to the time arrival of the reference pulse, which can be set by a sub-micrometric step delay line.
Each interferogram is analyzed by using Fast
Fourier Transform (FFT), in the reciprocal space,
where the off-axis term contains the modulation information associated to the emission at the given time.
This can be filtered to retrieve the dynamics of the polariton fluid, in both amplitude and phase.
The polariton phase maps are processed with a digital algorithm to
retrieve all the phase singularities, whose trajectories are rebuilt.
The used spatial and temporal steps here are $0.2064~\upmu\text{m}$ and $0.5~\text{ps}$, respectively,
while the time resolution is set by the reference pulse itself to $4\text{ ps}$.
Every time-frame results from tens of thousands of repeated shots, which are
integrated by the CCD camera set in a range between $0.15\text{ ms}$ and $1.0\text{ ms}$.
The visibility of the fringes remains stable for values $\tau_{\rm CCD} \leq$ 1.0 ms,
 which cuts out the mechanical vibrations of the setup.
This procedure allows to follow deterministic vortex trajectories,
while eventual stochastic (random path) vortices are washed
away by the averaging of the integration process.
\edstrike{Additional details on the technique can be found in Refs.~\cite{Dominici2014,dominici_vortex_2015}.}\\

\noindent \textbf{Theoretical model.}
In the polariton literature, there have been extensive
  studies in both the realms of incoherent pumping\edstrike{, based
  on the model of Ref.}~\cite{WC2007}\edstrike{,} and coherent coupling\edstrike{,
as in Refs.}~\cite{dominici_real-space_2015,dominici_vortex_2015}.
The present setting belongs to the latter kind, as we are resonantly pumping our sample.
Therefore,
to simulate our experimental setup we use a model of four coupled
Gross-Pitaevskii (GP) equations for the two spin (circular polarization)
components for both excitons and photons:
\begin{eqnarray}
\label{eq:coupled_GPEs_1}
\text{i}\hbar \frac{\partial \psi_{\pm} }{\partial t} &=&
\left(-\frac{\hbar^2}{2m_\psi}\nabla^2 - \text{i} \frac{\hbar}{2\tau_\psi} \right) \psi_\pm + \frac{\hbar\Omega_\text{R}}{2}\phi_{\pm}
\\
\notag
&&
 + g_{11}|\psi_\pm|^2 \psi_\pm + g_{12} |\psi_\mp|^2\psi_\pm,
\\
\text{i}\hbar \frac{\partial \phi_{\pm}}{\partial t} &=&
\left(-\frac{\hbar^2}{2m_\phi}\nabla^2 - \text{i} \frac{\hbar}{2\tau_\phi}
\right) \phi_\pm + \frac{\hbar\Omega_\text{R}}{2}\psi_{\pm}
\label{eq:coupled_GPEs_2}
\\
\notag
&&
+ \chi \left( \frac{\partial}{\partial x}\pm \text{i} \frac{\partial}{\partial
  y}\right)^2 \phi_{\mp} + F_{\pm},
\end{eqnarray}
where $\psi_\pm$ and $\phi_\pm$ represent, respectively, the wavefunctions
for the excitons and photons ($\pm$ indicates the two polarizations),
$m_\psi$ and $m_\phi$ are the respective masses and
$\tau_\psi$ and $\tau_\phi$ are the respective lifetimes,
$\Omega_\text{R}$ is the Rabi coupling frequency,
$g_{11}$ is the intra-spin exciton-exciton interaction strength while
$g_{12}$ represents the inter-spin interaction strength,
$\chi$ is the coefficient of the spin-orbit coupling,
and
$F_{\pm}$ is the applied external laser pulse.
The value for the Rabi splitting is taken to be $\Omega_\text{R}=5.4~\text{meV}$
and the exciton-exciton interaction strength~\cite{dominici_real-space_2015}
$g_{11}=2.0~\upmu\text{eV}\cdot\upmu\text{m}^2$.
The exciton and photon lifetimes relevant for this experiment are
$\tau_\psi=1000~\text{ps}$ and $\tau_\phi=5~\text{ps}$ respectively.
We consider the strength of the inter-spin exciton interaction to be an
order of magnitude weaker than the intra-spin interaction, so that
$g_{12}=-0.1 g_{11}$.
The photon mass $m_{\phi}=3.6\times 10^{-5}\,m_\text{e}$, where $m_\text{e}$ is the electron mass,
was extracted from the dispersion relationship.
On the other hand, the exciton mass is approximately four orders of
magnitude larger than the photonic one, so that the exciton kinetic
term could be in principle safely neglected.
Also, we treat the mass of the transverse electric component of the cavity mode,
$m_\phi^\text{TE}$ as being around $95\%$ that of the transverse
magnetic component $m_\phi^\text{TM}(\approx5\times10^{-5}m_\text{e})$, so
that the strength of the TE-TM splitting is taken to be $\chi=
\frac{\hbar^2}{4}(\frac{1}{m_\phi^\text{TE}}-\frac{1}{m_\phi^\text{TM}})
= 0.019 \times\frac{\hbar^2}{2m_\phi}$.
Finally, the external laser pulse is modelled as a coherent pump term in the photon field of Eq.~(\ref{eq:coupled_GPEs_2}), by writing
\begin{equation}
F_\pm(\mathbf{r},t) = f_\pm \times R_{\pm}(\mathbf{r}) \times T(t),
\end{equation}
where for the spatial part $R_{\pm}$ we used the normalized experimentally
measured 2D spatial profiles---corresponding to a modified Laguerre-Gauss profile with the appropriate
number of vortices that are \edstrike{imprinted} \refer{seeded} in the condensate---and for the temporal part $T(t)$ we used
\begin{equation}
T(t) = \text{e}^{- \frac{\left(t-t_\text{0}\right)^2}{2\sigma_t^2} }.
\end{equation}
The strength of the laser pulse, $f_\pm$, is chosen so as to
replicate the observed total photon output in the experiments.
The duration of the probe $\sigma_t$ was chosen in order
to correspond to a $\text{FWHM}_t=4\text{ ps}$, in line
with the experimental realization.
The pump is instantiated some time
into the simulation, reaching its maximum at $t_\text{0}=5.5\text{ ps}$
and removed completely after $5\sigma_t$ so as to negate any
unintended phase-locking.\\

\noindent \editr{
\textbf{Data availability.} All the original interferograms 
produced from the experiments of this study 
are available upon request from the corresponding author.\\
}

}

\bigskip

\noindent \textbf{Acknowledgments}\\
We thank Romuald Houdr\'{e} and
Alberto Bramati for the microcavity sample.
We kindly acknowledge Lorenzo Marrucci and Bruno Piccirillo for providing the $q$-plate devices.
We acknowledge the European Research Council project POLAFLOW (Grant 308136),
the Italian Ministero dell'Istruzione dell'Universit\'{a} e della Ricerca project ``Beyond Nano''
and
the project ``Molecular nAnotechnologies for heAlth and environmenT''
(MAAT, PON02-00563-3316357 and CUP B31C12001230005)
for financial support.
R.C.G.~and P.G.K.~acknowledge support from NSF-DMS-1309035, PHY-1603058,
NSF-DMS-1312856, and PHY-1602994.
J.C.M.~thanks financial support from MAT2016-79866-R project (AEI/FEDER, UE).\\

\bigskip

\def\bibsection{\section*{\refname}} 


\begin{thebibliography}{10}

\bibitem{abbott_observation_2016}
{\sc B.~P. Abbott and {LIGO Scientific Collaboration and Virgo Collaboration}},
  {\em Observation of {Gravitational} {Waves} from a {Binary} {Black} {Hole}
  {Merger}}, Phys. Rev. Lett., 116 (2016), p.~061102.

\bibitem{Amo2009}
{\sc A.~Amo, J.~Lefr\`{e}re, S.~Pigeon, C.~Adrados, C.~Ciuti, I.~Carusotto,
  R.~Houdr\'{e}, E.~Giacobino, and A.~Bramati}, {\em {Superfluidity of
  polaritons in semiconductor microcavities}}, Nat. Phys., 5 (2009),
  pp.~805--810.

\bibitem{Amo2011}
{\sc A.~Amo, S.~Pigeon, D.~Sanvitto, V.~G. Sala, R.~Hivet, I.~Carusotto,
  F.~Pisanello, G.~Lem\'{e}nager, R.~Houdr\'{e}, E.~Giacobino, C.~Ciuti, and
  A.~Bramati}, {\em {Polariton superfluids reveal quantum hydrodynamic
  solitons.}}, Science, 332 (2011), pp.~1167--1170.

\bibitem{Anton2012}
{\sc C.~Ant\'{o}n, G.~Tosi, M.~D. Mart\'{\i}n, L.~Vi\~{n}a, A.~Lema\^{\i}tre,
  and J.~Bloch}, {\em {Role of supercurrents on vortices formation in polariton
  condensates}}, Opt. Express, 20 (2012), pp.~16366--16373.

\bibitem{blatter_vortices_1994}
{\sc G.~Blatter, M.~V. Feigel'man, V.~B. Geshkenbein, A.~I. Larkin, and V.~M.
  Vinokur}, {\em Vortices in high-temperature superconductors}, Rev. Mod.
  Phys., 66 (1994), pp.~1125--1388.

\bibitem{boulier_vortex_2015}
{\sc T.~Boulier, H.~Ter\c{c}as, D.~D. Solnyshkov, Q.~Glorieux, E.~Giacobino,
  G.~Malpuech, and A.~Bramati}, {\em {Vortex Chain in a Resonantly Pumped
  Polariton Superfluid}}, Sci. Rep., 5 (2015), p.~9230.

\bibitem{Byrnes2014}
{\sc T.~Byrnes, N.~Y. Kim, and Y.~Yamamoto}, {\em Exciton-polariton
  condensates}, Nat. Phys., 10 (2014), pp.~803--813.

\bibitem{calderaro_vortex_2017}
{\sc L.~Calderaro, A.~L. Fetter, P.~Massignan, and P.~Wittek}, {\em Vortex
  dynamics in coherently coupled {Bose}-{Einstein} condensates}, Phys. Rev. A,
  95 (2017), p.~023605.

\bibitem{caputo_topological_2017}
{\sc D.~Caputo, D.~Ballarini, G.~Dagvadorj, C.~S\'{a}nchez Mu\~{n}oz,
  M.~De Giorgi, L.~Dominici, K.~West, L.~N. Pfeiffer, G.~Gigli, F.~P. Laussy,
  M.~Szyma\'{n}ska, and D.~Sanvitto}, {\em Topological order and thermal
  equilibrium in polariton condensates}, Nat. Mater., 17 (2017), pp.~145--151.

\bibitem{Cardano2013}
{\sc F.~Cardano, E.~Karimi, L.~Marrucci, C.~de~Lisio, and E.~Santamato}, {\em
  {Generation and dynamics of optical beams with polarization singularities.}},
  Opt. Express, 21 (2013), pp.~8815--8820.

\bibitem{dagvadorj_nonequilibrium_2015}
{\sc G.~Dagvadorj, J.~M. Fellows, S.~Matyja\'{s}kiewicz, F.~M. Marchetti,
  I.~Carusotto, and M.~H. Szyma\'{n}ska}, {\em Nonequilibrium {Phase}
  {Transition} in a {Two}-{Dimensional} {Driven} {Open} {Quantum} {System}},
  Phys. Rev. X, 5 (2015), p.~041028.

\bibitem{Dominici2014}
{\sc L.~Dominici, D.~Colas, S.~Donati, J.~P. Restrepo~Cuartas, M.~De~Giorgi,
  D.~Ballarini, G.~Guirales, J.~C. L\'{o}pez~Carre\'{n}o, A.~Bramati, G.~Gigli,
  E.~del Valle, F.~P. Laussy, and D.~Sanvitto}, {\em Ultrafast control and
  {Rabi} oscillations of polaritons}, Phys. Rev. Lett., 113 (2014), p.~226401.

\bibitem{dominici_ultrafast_2018}
{\sc L.~Dominici, D.~Colas, A.~Gianfrate, A.~Rahmani, C.~S. Muñoz,
  D.~Ballarini, M.~De~Giorgi, G.~Gigli, F.~P. Laussy, and D.~Sanvitto}, {\em
  Ultrafast topology shaping by a {Rabi}-oscillating vortex.},  (2018),
  p.~Preprint at http://arxiv.org/abs/1801.02580.

\bibitem{dominici_vortex_2015}
{\sc L.~Dominici, G.~Dagvadorj, J.~M. Fellows, D.~Ballarini, M.~D. Giorgi,
  F.~M. Marchetti, B.~Piccirillo, L.~Marrucci, A.~Bramati, G.~Gigli, M.~H.
  Szyma{\'n}ska, and D.~Sanvitto}, {\em Vortex and half-vortex dynamics in a
  nonlinear spinor quantum fluid}, Sci. Adv., 1 (2015), p.~e1500807.

\bibitem{dominici_real-space_2015}
{\sc L.~Dominici, M.~Petrov, M.~Matuszewski, D.~Ballarini, M.~De~Giorgi,
  D.~Colas, E.~Cancellieri, B.~Silva~Fern\'{a}ndez, A.~Bramati, G.~Gigli,
  A.~Kavokin, F.~Laussy, and D.~Sanvitto}, {\em Real-space collapse of a
  polariton condensate}, Nat. Commun., 6 (2015), p.~8993.

\bibitem{donati_twist_2016}
{\sc S.~Donati, L.~Dominici, G.~Dagvadorj, D.~Ballarini, M.~De~Giorgi,
  A.~Bramati, G.~Gigli, Y.~G. Rubo, M.~H. Szyma\'{n}ska, and D.~Sanvitto}, {\em
  Twist of generalized skyrmions and spin vortices in a polariton superfluid},
  Proc. Natl. Acad. Sci., 113 (2016), pp.~14926--14931.

\bibitem{eto_interaction_2011}
{\sc M.~Eto, K.~Kasamatsu, M.~Nitta, H.~Takeuchi, and M.~Tsubota}, {\em
  Interaction of half-quantized vortices in two-component {Bose}-{Einstein}
  condensates}, Phys. Rev. A, 83 (2011), p.~063603.

\bibitem{fedi_superfluid_2016}
{\sc M.~Fedi}, {\em {A superfluid Theory of Everything?}},  (2016),
  pp.~Preprint at https://hal.archives--ouvertes.fr/hal--01312579.

\bibitem{fetter}
{\sc A.~L. Fetter}, {\em Rotating trapped {Bose-Einstein} condensates}, Rev.
  Mod. Phys., 81 (2009), pp.~647--691.

\bibitem{Franchetti2012}
{\sc G.~Franchetti, N.~G. Berloff, and J.~J. Baumberg}, {\em {Exploiting
  quantum coherence of polaritons for ultra sensitive detectors.}},  (2012),
  p.~Preprint at https://arxiv.org/abs/1210.1187.

\bibitem{roumpos2}
{\sc M.~D. Fraser, G.~Roumpos, and Y.~Yamamoto}, {\em Vortex-antivortex pair
  dynamics in an exciton-polariton condensate}, New J. Phys., 11 (2009),
  p.~113048.

\bibitem{Freilich1182}
{\sc D.~V. Freilich, D.~M. Bianchi, A.~M. Kaufman, T.~K. Langin, and D.~S.
  Hall}, {\em Real-time dynamics of single vortex lines and vortex dipoles in a
  {Bose-Einstein} condensate}, Science, 329 (2010), pp.~1182--1185.

\bibitem{gianfrate_superluminal_2018}
{\sc A.~Gianfrate, L.~Dominici, O.~Voronych, M.~Matuszewski, M.~Stobi\'{n}ska,
  D.~Ballarini, M.~De~Giorgi, G.~Gigli, and D.~Sanvitto}, {\em Superluminal
  {X-waves} in a polariton quantum fluid}, Light Sci. Appl., 7 (2018),
  p.~e17119.

\bibitem{Hivet2014}
{\sc R.~Hivet, E.~Cancellieri, T.~Boulier, D.~Ballarini, D.~Sanvitto, F.~M.
  Marchetti, M.~H. Szymanska, C.~Ciuti, E.~Giacobino, and A.~Bramati}, {\em
  Interaction-shaped vortex-antivortex lattices in polariton fluids}, Phys.
  Rev. B, 89 (2014), p.~134501.

\bibitem{huang_quantum_2015}
{\sc K.~Huang}, {\em Quantum vorticity in nature}, Int. J. Mod. Phys. A, 30
  (2015), p.~1530056.

\bibitem{Huang20161}
{\sc K.~Huang}, {\em A superfluid universe}, World Scientific Publishing Co.
  Pte. Ltd., 2016.

\bibitem{Jackson1998}
{\sc B.~Jackson, J.~F. McCann, and C.~S. Adams}, {\em Vortex formation in
  dilute inhomogeneous {Bose-Einstein} condensates}, Phys. Rev. Lett., 80
  (1998), pp.~3903--3906.

\bibitem{kasamatsu_short-range_2016}
{\sc K.~Kasamatsu, M.~Eto, and M.~Nitta}, {\em Short-range intervortex
  interaction and interacting dynamics of half-quantized vortices in
  two-component {Bose}-{Einstein} condensates}, Phys. Rev. A, 93 (2016),
  p.~013615.

\bibitem{kasamatsu_vortex_2004}
{\sc K.~Kasamatsu, M.~Tsubota, and M.~Ueda}, {\em Vortex {Molecules} in
  {Coherently} {Coupled} {Two}-{Component} {Bose}-{Einstein} {Condensates}},
  Phys. Rev. Lett., 93 (2004), p.~250406.

\bibitem{kasamatsu_multi_review_2005}
\leavevmode\vrule height 2pt depth -1.6pt width 23pt, {\em Vortices in
  multicomponent {B}ose-{E}instein condensates}, Int. J. Mod. Phys. B, 19
  (2005), pp.~1835--1904.

\bibitem{Kasprzak2006}
{\sc J.~Kasprzak, M.~Richard, S.~Kundermann, A.~Baas, P.~Jeambrun, J.~M.~J.
  Keeling, F.~M. Marchetti, M.~H. Szyma\'{n}ska, R.~Andr\'{e}, J.~L. Staehli,
  V.~Savona, P.~B. Littlewood, B.~Deveaud, and L.~S. Dang}, {\em {Bose-Einstein
  condensation of exciton polaritons.}}, Nature, 443 (2006), pp.~409--414.

\bibitem{Kevrekidis2016140}
{\sc P.~Kevrekidis and D.~Frantzeskakis}, {\em Solitons in coupled nonlinear
  {Schr{\"o}dinger} models: A survey of recent developments}, Rev. Phys., 1
  (2016), pp.~140 -- 153.

\bibitem{Kevrekidis_MPLB_2004}
{\sc P.~G. Kevrekidis, R.~Carretero-Gonz\'alez, D.~J. Frantzeskakis, and I.~G.
  Kevrekidis}, {\em Vortices in {B}ose-{E}instein condensates: Some recent
  developments}, Mod. Phys. Lett. B, 18 (2004), pp.~1481--1505.

\bibitem{DarkBook}
{\sc P.~G. Kevrekidis, D.~J. Frantzeskakis, and R.~Carretero-Gonz\'{a}lez},
  {\em The defocusing nonlinear Schr{\"o}dinger equation: from dark solitons
  and vortices to vortex rings}, SIAM, Philadelphia, 2015.

\bibitem{kivshar_dynamics_1998}
{\sc Y.~S. Kivshar, J.~Christou, V.~Tikhonenko, B.~Luther-Davies, and L.~M.
  Pismen}, {\em Dynamics of optical vortex solitons}, Opt. Commun., 152 (1998),
  pp.~198--206.

\bibitem{theo}
{\sc T.~Kolokolnikov, P.~G. Kevrekidis, and R.~Carretero-Gonz{\'a}lez}, {\em A
  tale of two distributions: from few to many vortices in quasi-two-dimensional
  {Bose-Einstein} condensates}, Proc. R. Soc. Lond. A, 470 (2014), p.~20140048.

\bibitem{Lagoudakis2009}
{\sc K.~G. Lagoudakis, T.~Ostatnick\'{y}, A.~V. Kavokin, Y.~G. Rubo,
  R.~Andr\'{e}, and B.~Deveaud-Pl\'{e}dran}, {\em Observation of half-quantum
  vortices in an exciton-polariton condensate}, Science, 326 (2009),
  pp.~974--976.

\bibitem{Lagoudakis2008}
{\sc K.~G. Lagoudakis, M.~Wouters, M.~Richard, A.~Baas, I.~Carusotto,
  R.~Andr\'{e}, L.~S. Dang, and B.~Deveaud-Pl\'{e}dran}, {\em Quantized
  vortices in an exciton-polariton condensate}, Nat. Phys., 4 (2008),
  pp.~706--710.

\bibitem{li_dynamics_2016}
{\sc T.~Li, S.~Yi, and Y.~Zhang}, {\em Dynamics of a coupled spin-vortex pair
  in dipolar spinor {Bose}-{Einstein} condensates}, Phys. Rev. A, 93 (2016),
  p.~053602.

\bibitem{Liew2007}
{\sc T.~C.~H. Liew, A.~V. Kavokin, and I.~A. Shelykh}, {\em Excitation of
  vortices in semiconductor microcavities}, Phys. Rev. B, 75 (2007),
  p.~241301(R).

\bibitem{Liu2015}
{\sc G.~Liu, D.~W. Snoke, A.~Daley, L.~N. Pfeiffer, and K.~West}, {\em A new
  type of half-quantum circulation in a macroscopic polariton spinor ring
  condensate}, Proc. Natl. Acad. Sci., 112 (2015), pp.~2676--2681.

\bibitem{Manni2012}
{\sc F.~Manni, K.~Lagoudakis, and T.~C.~H. Liew}, {\em {Dissociation dynamics
  of singly charged vortices into half-quantum vortex pairs}}, Nat. Commun., 3
  (2012), p.~1309.

\bibitem{Manni2013}
{\sc F.~Manni, Y.~L\'{e}ger, Y.~G. Rubo, R.~Andr\'{e}, and B.~Deveaud}, {\em
  {Hyperbolic spin vortices and textures in exciton-polariton condensates}},
  Nat. Commun., 4 (2013), p.~2590.

\bibitem{Marrucci2006}
{\sc L.~Marrucci, C.~Manzo, and D.~Paparo}, {\em {Optical Spin-to-Orbital
  Angular Momentum Conversion in Inhomogeneous Anisotropic Media}}, Phys. Rev.
  Lett., 96 (2006), p.~163905.

\bibitem{middelkamp_guiding-center_2011}
{\sc S.~Middelkamp, P.~J. Torres, P.~G. Kevrekidis, D.~J. Frantzeskakis,
  R.~Carretero-Gonz\'{a}lez, P.~Schmelcher, D.~V. Freilich, and D.~S. Hall},
  {\em Guiding-center dynamics of vortex dipoles in {Bose}-{Einstein}
  condensates}, Phys. Rev. A, 84 (2011), p.~011605(R).

\bibitem{molina-terriza_twisted_2007}
{\sc G.~Molina-Terriza, J.~P. Torres, and L.~Torner}, {\em Twisted photons},
  Nat. Phys., 3 (2007), pp.~305--310.

\bibitem{Nardin2010}
{\sc G.~Nardin, K.~G. Lagoudakis, B.~Pietka, F.~Morier-Genoud, Y.~L\'{e}ger,
  and B.~Deveaud-Pl\'{e}dran}, {\em {Selective photoexcitation of confined
  exciton-polariton vortices}}, Phys. Rev. B, 82 (2010), p.~073303.

\bibitem{navarro_dynamics_2013}
{\sc R.~Navarro, R.~Carretero-Gonz{\'a}lez, P.~J. Torres, P.~G. Kevrekidis,
  D.~J. Frantzeskakis, M.~W. Ray, E.~Altunta{\c{s}}, and D.~S. Hall}, {\em
  Dynamics of a {Few} {Corotating} {Vortices} in {Bose}-{Einstein}
  {Condensates}}, Phys. Rev. Lett., 110 (2013), p.~225301.

\bibitem{BPA}
{\sc T.~W. Neely, A.~S. Bradley, E.~C. Samson, S.~J. Rooney, E.~M. Wright,
  K.~J.~H. Law, R.~Carretero-Gonz\'alez, P.~G. Kevrekidis, M.~J. Davis, and
  B.~P. Anderson}, {\em Characteristics of two-dimensional quantum turbulence
  in a compressible superfluid}, Phys. Rev. Lett., 111 (2013), p.~235301.

\bibitem{nitta_vortex_2014}
{\sc M.~Nitta, M.~Eto, and M.~Cipriani}, {\em Vortex {Molecules} in
  {Bose}-{Einstein} {Condensates}}, J. Low Temp. Phys., 175 (2014),
  pp.~177--188.

\bibitem{stringari}
{\sc L.~Pitaevskii and S.~Stringari}, {\em Bose-Einstein Condensation}, Oxford
  University Press, Oxford, 2003.

\bibitem{pshenichnyuk_pair_2017}
{\sc I.~A. Pshenichnyuk}, {\em Pair interactions of heavy vortices in quantum
  fluids.},  (2017), p.~Preprint at http://arxiv.org/abs/1705.10072.

\bibitem{Rodrigues2014}
{\sc A.~S. Rodrigues, P.~G. Kevrekidis, R.~Carretero-Gonz\'{a}lez,
  J.~Cuevas-Maraver, D.~J. Frantzeskakis, and F.~Palmero}, {\em From nodeless
  clouds and vortices to gray ring solitons and symmetry-broken states in
  two-dimensional polariton condensates}, J. Phys.: Condens. Matter, 26 (2014),
  p.~155801.

\bibitem{Rubo2007}
{\sc Y.~Rubo}, {\em {Half vortices in exciton polariton condensates}}, Phys.
  Rev. Lett., 99 (2007), p.~106401.

\bibitem{sanvitto_road_2016}
{\sc D.~Sanvitto and S.~K\'{e}na-Cohen}, {\em The road towards polaritonic
  devices}, Nat. Mater., 15 (2016), pp.~1061--1073.

\bibitem{Sanvitto2010}
{\sc D.~Sanvitto, F.~M. Marchetti, M.~H. Szyma\'{n}ska, G.~Tosi, M.~Baudisch,
  F.~P. Laussy, D.~N. Krizhanovskii, M.~S. Skolnick, L.~Marrucci,
  A.~Lema\^{\i}tre, J.~Bloch, C.~Tejedor, and L.~Vi\~{n}a}, {\em {Persistent
  currents and quantized vortices in a polariton superfluid}}, Nat. Phys., 6
  (2010), pp.~527--533.

\bibitem{sanvitto_all-optical_2011}
{\sc D.~Sanvitto, S.~Pigeon, A.~Amo, D.~Ballarini, M.~De~Giorgi, I.~Carusotto,
  R.~Hivet, F.~Pisanello, V.~G. Sala, P.~S.~S. Guimaraes, R.~Houdr\'e,
  E.~Giacobino, C.~Ciuti, A.~Bramati, and G.~Gigli}, {\em All-optical control
  of the quantum flow of a polariton condensate}, Nat. Photon., 5 (2011),
  pp.~610--614.

\bibitem{sbitnev_hydrodynamics_2016_I}
{\sc V.~I. Sbitnev}, {\em Hydrodynamics of the {Physical} {Vacuum}: {I}.
  {Scalar} {Quantum} {Sector}}, Found. Phys., 46 (2016), pp.~606--619.

\bibitem{sbitnev_hydrodynamics_2016_II}
\leavevmode\vrule height 2pt depth -1.6pt width 23pt, {\em Hydrodynamics of the
  {Physical} {Vacuum}: {II}. {Vorticity} {Dynamics}}, Found. Phys., 46 (2016),
  pp.~1238--1252.

\bibitem{Schnars_book2005a}
{\sc U.~Schnars and W.~J\"{u}ptner}, {\em {Digital Holography}}, Springer
  Berlin Heidelberg, 2005.

\bibitem{seo_collisional_2016}
{\sc S.~W. Seo, W.~J. Kwon, S.~Kang, and Y.~Shin}, {\em Collisional {Dynamics}
  of {Half}-{Quantum} {Vortices} in a {Spinor} {Bose}-{Einstein} {Condensate}},
  Phys. Rev. Lett., 116 (2016), p.~185301.

\bibitem{serafini_vortex_2017}
{\sc S.~Serafini, L.~Galantucci, E.~Iseni, T.~Bienaim\'e, R.~N. Bisset, C.~F.
  Barenghi, F.~Dalfovo, G.~Lamporesi, and G.~Ferrari}, {\em Vortex
  reconnections and rebounds in trapped atomic {Bose-Einstein} condensates},
  Phys. Rev. X, 7 (2017), p.~021031.

\bibitem{Sigurdsson2014}
{\sc H.~Sigurdsson, O.~A. Egorov, X.~Ma, I.~A. Shelykh, and T.~C.~H. Liew},
  {\em Information processing with topologically protected vortex memories in
  exciton-polariton condensates}, Phys. Rev. B, 90 (2014), p.~014504.

\bibitem{jeff}
{\sc J.~Steinhauer}, {\em Observation of quantum {Hawking} radiation and its
  entanglement in an analogue black hole}, Nat. Phys., 12 (2015), pp.~959--965.

\bibitem{torres_dynamics_2011}
{\sc P.~Torres, P.~Kevrekidis, D.~Frantzeskakis, R.~Carretero-Gonz{\'a}lez,
  P.~Schmelcher, and D.~Hall}, {\em Dynamics of vortex dipoles in confined
  {Bose}-{Einstein} condensates}, Phys. Lett. A, 375 (2011), pp.~3044--3050.

\bibitem{tsubota_quantum_2013}
{\sc M.~Tsubota, M.~Kobayashi, and H.~Takeuchi}, {\em Quantum hydrodynamics},
  Phys. Rep., 522 (2013), pp.~191--238.

\bibitem{tylutki_confinement_2016}
{\sc M.~Tylutki, L.~P. Pitaevskii, A.~Recati, and S.~Stringari}, {\em
  Confinement and precession of vortex pairs in coherently coupled
  {Bose}-{Einstein} condensates}, Phys. Rev. A, 93 (2016), p.~043623.

\bibitem{uchida_generation_2010}
{\sc M.~Uchida and A.~Tonomura}, {\em Generation of electron beams carrying
  orbital angular momentum}, Nature, 464 (2010), pp.~737--739.

\bibitem{Voronova2012}
{\sc N.~S. Voronova and Y.~E. Lozovik}, {\em Excitons in cores of
  exciton-polariton vortices}, Phys. Rev. B, 86 (2012), p.~195305.

\bibitem{willner_different_2012}
{\sc A.~E. Willner, J.~Wang, and H.~Huang}, {\em A {Different} {Angle} on
  {Light} {Communications}}, Science, 337 (2012), pp.~655--656.

\bibitem{WC2007}
{\sc M.~Wouters and I.~Carusotto}, {\em Excitations in a nonequilibrium
  {Bose-Einstein} condensate of exciton polaritons}, Phys. Rev. Lett., 99
  (2007), p.~140402.

\bibitem{zhao_pattern_2017}
{\sc H.~J. Zhao, V.~R. Misko, J.~Tempere, and F.~Nori}, {\em Pattern formation
  in vortex matter with pinning and frustrated intervortex interactions}, Phys.
  Rev. B, 95 (2017), p.~104519.

\bibitem{Zurek1985}
{\sc W.~H. Zurek}, {\em {Cosmological experiments in superfluid helium?}},
  Nature, 317 (1985), pp.~505--508.

\end{thebibliography}

\end{document}